\documentclass[12pt]{article}
\usepackage[utf8]{inputenc}
\usepackage{amsfonts,amsmath,amssymb}
\usepackage{braket}
\usepackage{graphicx}
\usepackage{color}
\begin{document}

\begin{titlepage}
\thispagestyle{empty}

\vspace*{-2cm}
\begin{flushright}
YITP-23-33, IPMU23-0014
\end{flushright}
\bigskip

\begin{center}
\noindent{{\large \textbf{Soft hair, dressed coordinates and information loss paradox}}}\\
\vspace{1cm}
Francesco Di Filippo$^{a}$, Naoki Ogawa$^{a}$, Shinji Mukohyama$^{a,b}$, Takahiro Waki$^{a}$
\vspace{1cm}\\
{\it $^a$Center for Gravitational Physics and Quantum Information,\\
Yukawa Institute for Theoretical Physics, Kyoto University, \\
Kitashirakawa Oiwakecho, Sakyo-ku, Kyoto 606-8502, Japan\\
$^b$Kavli Institute for the Physics and Mathematics of the Universe (WPI),\\
The University of Tokyo Institutes for Advanced Study (UTIAS),\\
The University of Tokyo, Kashiwa, Chiba 277-8583, Japan
}
\end{center}
\bigskip \bigskip

\begin{abstract}
Understanding the dynamics of soft hair might shine some light on the information loss paradox. In this paper, we introduce a new coordinate system, dressed coordinates, in order to analyze the quantum states of Hawking radiation as a first step toward understanding the connection between soft hair and the  information paradox. Dressed coordinates can be introduced by an operator-dependent coordinate transformation that makes the soft hair degrees of freedom apparently disappear from the metric. We show that some results of previous studies, such as the angle-dependent Hawking temperature and the soft graviton theorem, can be easily reproduced using these dressed coordinates. Finally, we discuss future possible applications of dressed coordinates toward understanding the information loss paradox.
\end{abstract}
\end{titlepage}
\clearpage

\section{Introduction}

The infrared (IR) triangle is a universal relation between three sectors of physics in the infrared of gravitational (or gauge) theory, the Bondi–Metzner–Sachs (BMS) symmetry, the memory effects, and the soft graviton theorem~\cite{Strominger:2013jfa,He:2014laa,Strominger:2014pwa,Strominger:2017zoo}. The BMS symmetry is a symmetry that preserves asymptotic structures at null infinity in asymptotically flat spacetimes and can be written as a semidirect product of supertranslation and Lorentz transformation~\cite{Bondi:1962px,Sachs:1962wk}. The supertranslation is an angle-dependent time translation at null infinity, which is a nontrivial symmetry of asymptotic regions.

The key consideration underlying this relationship is that supertranslated spacetimes can be regarded as physically different spacetimes from the original. Supertranslated spacetimes are spacetimes with very low energy gravitons, soft gravitons. In classical theory, this induces memory effects, and in quantum theory, this leads to soft graviton theorems.

In the case of black hole spacetimes, we can obtain a new spacetime with soft gravitons by applying supertranslation to the original Schwarzschild spacetime without soft gravitons~\cite{Hawking:2016msc,Hawking:2016sgy}. This new degree of freedom for black holes due to soft gravitons is called soft hair. Since supertranslation is a diffeomorphism map, supertranslated spacetimes are locally equivalent to the original spacetimes. However, these diffeomorphic spacetimes are globally different from the original spacetime, thus avoiding the no-hair theorem. This feature is attracting attention as an approach to problems in black holes, such as the micro origin of entropy~\cite{Afshar:2016wfy,Haco:2018ske,Afshar:2016kjj,Afshar:2017okz} and the information paradox~\cite{Strominger:2017aeh,Pasterski:2020xvn,Hotta:2017yzk,Flanagan:2021ojq,Flanagan:2021svq}.

A possible clue to solve the information paradox with soft hair was proposed by Hawking \textit{et al.}~\cite{Hawking:2016msc,Hawking:2016sgy,Hawking:2015qqa}, where it is argued that the presence of infinite conserved quantities can constrain Hawking radiation. This possibility is currently under debate as opposite results have been reported. In Refs.~\cite{Bousso:2017dny,Mirbabayi:2016axw} the authors argue that the dynamics of hard particles (including Hawking quanta) cannot be influenced by soft gravitons. Hence, the conservation of asymptotic symmetry charges could only constrain soft particles. An opposite conclusion is reached in Ref.~\cite{Flanagan:2022pmj}, where a direct computation obtains a result that seems to contradict this statement. As a consequence, the possible relevance of soft hair in the resolution of the information loss paradox is unclear. As explained in~\cite{Flanagan:2022pmj}, it is possible that the result of \cite{Bousso:2017dny,Mirbabayi:2016axw} does not hold due to the use of non-independent coordinates to describe the phase space. On the other hand, to show that these coordinates are not independent, we must assume the continuity of some quantities when approaching the asymptotic boundary. In Ref.~\cite{Bousso:2017rsx} it is discussed that this continuity is not ensured~\footnote{We would like to thank Massimo Porrati for pointing out this issue to us.}.

Remaining, for the moment, agnostic to the resolution of this debate, with the final goal to investigate the interplay between soft hair and the information loss problem, in this paper we introduce the concept of ``dressed coordinates'', which are coordinates that depend on the degrees of freedom of the soft hair. 

In section~\ref{Sec:dressed_coord} we introduce the definition of dressed coordinates, and in section~\ref{REP} we show the relevance of such coordinates by providing an alternative derivation of well-known results using them. Let us emphasize that similar concepts are  also introduced in \cite{Bousso:2017dny,Mirbabayi:2016axw} in a slightly different way as we want to highlight the relevance of this concept, and we explicitly clarify that this is simply a choice of coordinates that makes the metric apparently soft graviton free. Furthermore, we differentiate between soft gravitons in the asymptotic region and horizon soft hair. 

In section \ref{HR}, we calculate Hawking radiation of soft hair black holes. The results are consistent with previous studies~\cite{Chu:2018tzu}. As expected, Hawking radiation without backreaction is not influenced by soft hair. The purpose of this work is to introduce dressed coordinates that significantly simplify the derivation of known results. Furthermore, we argue that in the future the dressed coordinates might allow the computation of Hawking radiation including backreaction.

In section \ref{DIS}, we discuss the expectation of the effect of soft hair on information paradox. We discuss what we could expect to happen if we include backreaction and how information needs to be transferred if the information loss problem is to be solved by soft hair. Finally, we suggest future directions.

\section{Soft hair and dressed coordinate}\label{Sec:dressed_coord}

In this section, we introduce the notion of dressed coordinates. As mentioned above, we distinguish between soft gravitons in asymptotic regions and horizon soft hairs. The former can be probed with physical quantities in the Hilbert space of the asymptotic region, while the latter can only be probed with physical quantities in the horizon Hilbert space. In this section, we clarify these distinctions and introduce a simple way to describe soft hair by introducing soft graviton dependent coordinates, the dressed coordinates.

\subsection{soft gravitons in asymptotic region}

\subsubsection{Soft graviton as Nambu-Goldstone boson}

We introduce the soft graviton degrees of freedom as Nambu-Goldstone (NG) bosons as Refs.~\cite{Strominger:2017zoo,Hawking:2016msc,Hawking:2016sgy}. In retarded Bondi coordinates, which are the natural coordinates used by observers in the null asymptotic future ${\mathcal{I}}^+$, the Schwarzschild spacetime metric without soft graviton can be written as follows:
\begin{equation} \label{eq:metric_Bondi}
g= -Vdu^2 - 2dudr + r^2\gamma_{AB}d\Theta^A\Theta^B\,,
\end{equation}
where $V=1-\frac{2m}{r}$. We introduce the degrees of freedom of the soft graviton in this metric. Let us recall that, in the retarded Bondi gauge, the vector generating supertranslation to ${\mathcal{I}}^+$ in Schwarzschild spacetime can be written as follows:
\begin{equation}
\zeta_f = f\partial_u +\frac{1}{r}D^Af\partial_A -\frac{1}{2}D^2f\partial_r\,.
\end{equation}
The radial and angular components are $\mathcal{O}(1/r)$ and are gauge-dependent. The retarded time component does not fall at infinity and is gauge independent. Using this vector, we can obtain a spacetime supertranslated with the parameter $f$ by performing the coordinate transformation $x_{new}^\mu+\zeta_f^\mu= x_{old}^\mu$, 
\begin{align} \label{eq:metric_hair}
\begin{aligned}
g&=-\left(V-\frac{mD^2f}{r^2}\right)du^2 - 2dudr -dvd\Theta^A D_A(2Vf+D^2f)\\
&~~~~~~~~+(r^2\gamma_{AB}+2rD_AD_Bf-r\gamma_{AB}D^2f)d\Theta^Ad\Theta^B\,.
\end{aligned}
\end{align}
Despite the fact that the two metrics \eqref{eq:metric_Bondi} and \eqref{eq:metric_hair} are related by coordinates transformation, we should regard these spacetimes as physically distinguishable as the latter contains a soft graviton degree of freedom, i.e. supertranslation should not be regarded as a redundancy of the description for physics~\cite{Strominger:2013jfa,He:2014laa,Strominger:2014pwa,Strominger:2017zoo}. Furthermore, the parameter $f$ characterizes the metric on which supertranslation symmetry is acted. It is broken on the Schwarzschild spacetime and there is an NG boson. In order to incorporate the degrees of freedom of this NG boson, it is necessary to promote the parameter of the supertranslation $f$ to a degree of freedom $\tilde{C}$. Thus, this is a coordinate of the phase space in classical theory and should be quantized in quantum theory. This NG boson $\tilde{C}$  corresponds to the degrees of freedom of soft gravitons.

Thus, we obtain the following metric containing the effects of soft hair.
\begin{align} \label{SR}
\begin{aligned}
\tilde{g}&=-\left(V-\frac{mD^2\tilde{C}}{r^2}\right)du^2 - 2dudr-D_A(2V\tilde{C}+D^2\tilde{C})dud\Theta^A
\\&~~~~~~~~+(r^2\gamma_{AB}+2rD_AD_B\tilde{C}-r\gamma_{AB}D^2\tilde{C})d\Theta^Ad\Theta^B\,.
\end{aligned}
\end{align}
We use the notation of attaching tildes for all soft graviton dependent quantities.

\subsubsection{Dressed coordinate}

We now introduce new soft graviton dependent coordinates, dressed coordinates, by defining $\tilde{x}^\mu=x^\mu+\tilde{C}$. Using these coordinates, the metric containing soft hairs \eqref{SR} is apparently the same as Schwarzschild spacetime without soft graviton degrees of freedom \eqref{eq:metric_Bondi}. 
\begin{equation}
\tilde{g}=-\tilde{V}d\tilde{u}^2 + 2d\tilde{u}d\tilde{r} + \tilde{r}^2\gamma_{AB}d\tilde{\Theta}^A\tilde{\Theta}^B\,,
\end{equation}
where $\tilde{V}=1-\frac{2m}{\tilde{r}}$. However, this metric should be distinguished from the Schwarzschild spacetime without soft graviton \eqref{eq:metric_Bondi}. In fact, while ordinary soft graviton independent coordinates commute with any physical quantity as they are constant in the phase space, dressed coordinates depend on non-commutative quantities such as the supertranslation charge. The introduction of the dressed coordinates physically means that moving the observer eliminates the ``distortion'' of spacetime caused by soft gravitons. This is equivalent to the fact that by momentum-dependent Lorentz boost, we can always stop a ball that is in a superposition state of momentum. 

Note that $\tilde{C}$ acts on the state $\ket{\Psi}_{soft}=\int \mathcal{D}{C} \Psi[C] \ket{C}_{soft}$ of the soft Hilbert space and $\ket{\Psi}$ represents a state of the frame from the passive point of view. This is equivalent to the active (ordinary) point of view that sees $\ket{\Psi}_{soft}$ as a soft graviton state. In the passive point of view, the coordinate $u$ indicates a different position depending on the state of the frame $\ket{\Psi}_{soft}$. However, coordinate $\tilde{u}$ indicate the same position independent of the state of frame $\ket{\Psi}_{soft}$. In this sense, dressed coordinates are more natural than normal coordinates. In the remaining of this paper, we provide examples that show the usefulness of dressing coordinates. Let us remark that similar concepts are also introduced in \cite{Bousso:2017dny,Mirbabayi:2016axw}~\footnote{Due to the notation, the sign of $\tilde{C}$ is different.}. We differ from these works as we highlight the relevance of this concept, and we explicitly clarify that this is simply a coordinate choice that makes the metric apparently soft graviton free.

So far, we have discussed dressed coordinates associated with soft gravitons in ${\mathcal{I}}^+$. Soft gravitons in ${\mathcal{I}}^-$ and the corresponding dressed coordinates can also be introduced in a similar way.

\subsection{Horizon soft hair}

In this subsection, we introduce horizon soft hair~\cite{Hawking:2016msc,Hawking:2016sgy,Donnay:2015abr}. Horizon soft hair can be regarded as a degree of freedom for perturbation by soft gravitons at the event horizon. This degree of freedom can be introduced in the same way as ${\mathcal{I}}^+$.

First, we write the metric without soft hair using horizon penetrating coordinates 
\begin{equation}
g = -V dv_H^2+2dv_Hd\rho_H + r^2\gamma_{AB}d\Theta_H^A\Theta_H^B\,.
\end{equation}
These coordinates are convenient for discussing quantum theory from the point of view of an observer passing through the horizon. The subscript $H$ is intended to make it clear that this is different from the asymptotic observer's coordinate $v$. Physically, this is a reflection of the independence of the motion of the observer on the horizon and the observer in the asymptotic region.

When horizon supertranslation acts on the metric, we obtain the following:
\begin{equation}
g=-V dv_H^2+2dv_Hd\rho_H + r^2\gamma_{AB}d\Theta_H^A\Theta_H^B+(f~\text{term})\,.
\end{equation}
Consider this $f$ as a degree of freedom, and further promote it to the horizon soft hair operator $\hat{C}_H$ by quantum soft hair. We use the notation of attaching a hat for each horizon soft hair dependent quantity.
\begin{equation}
g=-Vdv_H^2+2dv_Hd\rho_H + r^2\gamma_{AB}d\Theta_H^A\Theta_H^B+(\hat{C}_H~\text{term})\,.
\end{equation}
Finally, we introduce dressed coordinates by absorbing soft graviton modes $\hat{C}_H$ into coordinates as well as soft gravitons in the asymptotic region case.
\begin{equation}
g=-\hat{V}d\hat{v}_H^2+2d\hat{v}_Hd\hat{\rho}_H + \hat{r}^2\hat{\gamma}_{AB}d\hat{\Theta}_H^A\hat{\Theta}_H^B\,.
\end{equation}
Now we note that $\hat{C}_H$ term vanishes because it is nothing more than an inverse transformation. The states in this can be determined by free-falling observers passing through the horizon.

\section{Reproducing previous researches using dressed coordinate }\label{REP}

\subsection{Angle dependence of Hawking temperature}

A previous work~\cite{Chu:2018tzu} shows that the temperature of an evaporating classical black hole with soft hair is angle dependent. To reproduce this result, we start considering a Vaidya spacetime
\begin{equation}
g=-Vdv_H^2+2dv_Hdr_H+r_H^2\gamma_{AB}d\Theta^Ad\Theta^B\,,
\end{equation}
where $V=1-\frac{2M(v_H)}{r}$ and $\gamma_{AB}d\Theta^Ad\Theta^B$ is the metric of the unit $2$-sphere. We act with a horizon supertranslation on this metric. Namely we have to include hat in all quantities in our notation. 
\begin{equation}
\hat{g}=-\hat{V}d\hat{v}_H^2+2d\hat{v}_Hd\hat{r}_H+\hat{r}_H^2\hat{\gamma}_{AB}d\hat{\Theta}^Ad\hat{\Theta}^B\,. 
\end{equation}
In this subsection, we provide a simple derivation of the result of \cite{Chu:2018tzu} assuming that the dynamical evolution is adiabatic. If we think about classical soft hair, the coordinates are classical quantities. In the adiabatic approximation, the temperature is given by the same expression of the static case in terms of the surface gravity~\cite{Barcelo:2010pj}, as 
\begin{equation}
\hat{T}=\frac{\hat{\kappa}(\hat{v}_H)}{2\pi}\,,
\end{equation}
where following \cite{Hayward:1997jp}, the surface gravity $\hat{\kappa}$ is defined using the Kodama vector $\hat{K}^{\mu}=\delta_{\hat{v}_H}^{\mu}$ as $\hat{K}^{\mu}\nabla_{[\nu}\hat{K}_{\mu]}=-\hat{\kappa}\hat{K}_{\nu}$. Obviously, this Hawking temperature in dressed coordinates depends on time, but it is spherically symmetric. 

We now substitute $\hat{v}_H=v_H+\hat{C}_H$ to the expression of the Hawking temperature and expand it up to the first order in $\hat{C}_H$.  We then obtain the following expression.
\begin{equation}
\hat{T}=\frac{1}{8\pi M(v_H)}\left(1-\hat{C}_H\frac{M^\prime(v_H)}{M(v_H)}\right)\,.
\end{equation}
This is the temperature of the apparent horizon as seen by observers at infinity, which depends on the angle. This is because this coordinate is not synchronized, so observers in the coordinate $v_H$ observe the temperature at ``different time'' $\hat{v}_H$ on the horizon.

\subsection{Ward identity}

In this subsection; we discuss a Ward identity already introduced in Refs.~\cite{Strominger:2013jfa,He:2014laa} that can be obtained in a different way using dressed coordinates. Let us consider a scattering amplitude of $n$ particles to $m$ particles in position space represented as a state of a frame $\ket{C=0}$, where $C$ is a parameter of supertranslated spacetimes as follows:
\begin{align}
\mathcal{A}=&\bra{C=0}_{soft}\otimes\bra{0}_{hard}\phi(u_1,z_1)\cdots\phi(u_n,z_n)\nonumber\\
&\times \mathcal{S}\phi(v_{1},z_{n+1})\cdots\phi(v_m,z_{n+m})\ket{0}_{hard}\otimes\ket{C=0}_{soft}\,,\label{WA1}
\end{align}
where $\phi(x)$ is local operator, $\ket{0}$ is vacuum of the hard Hilbert space and $\ket{C=0}$ is a state of the soft Hilbert space such that $\tilde{C}\ket{C=0}=0$. In addition, we use the fact that the soft modes of initial and final states are identical because of antipodal frame matching conditions~\cite{Strominger:2013jfa}.
\begin{equation}
    \left.C\right|_{\mathcal{I}^+_-}=\left.C\right|_{\mathcal{I}^-_+}\,.
\end{equation}

\begin{figure}\label{fig:tsca}
  \centering
  \includegraphics[width=15cm]{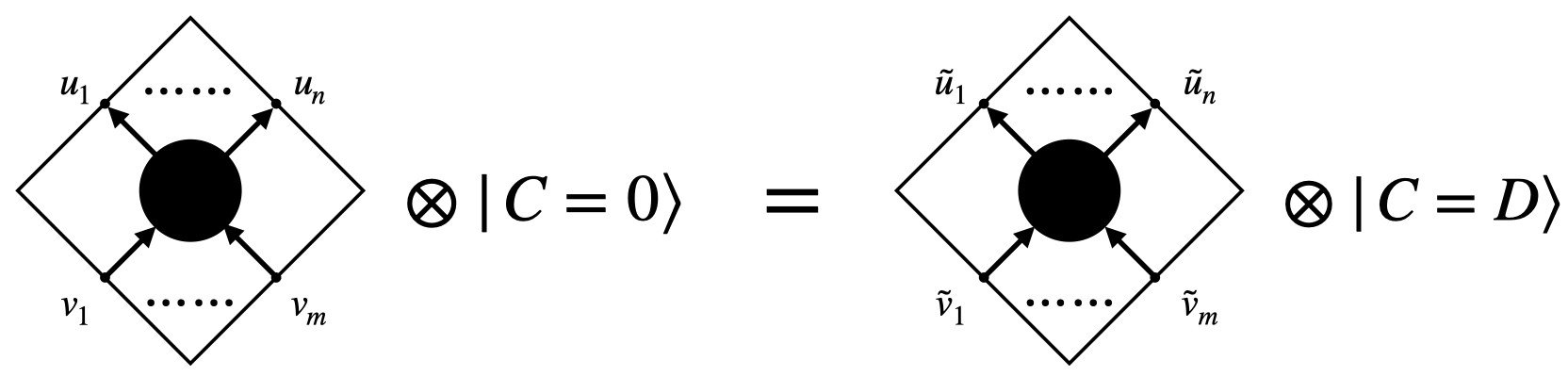}
  \caption{The Ward identity that relates scattering amplitudes before and after performing a supertranslation.}
\end{figure}

In a frame supertranslated with the parameter $C=D$, the above amplitude \eqref{WA1} can be written as follows.
\begin{align}
\mathcal{A}_{D}=&\bra{C=D}_{soft}\otimes\bra{0}_{hard}\phi(\tilde{u}_1,{z}_1)\cdots\phi(\tilde{u}_n,{z}_n)\nonumber\\
& \times\mathcal{S}\phi(\tilde{v}_{1},{z}_{n+1})\cdots\phi(\tilde{v}_m,{z}_{n+m})\ket{0}_{hard}\otimes\ket{C=D}_{soft}\,.
\label{titi}
\end{align}
From the discussion in the previous section, it is obvious that the position of $u$ in the frame of $\ket{C=0}_{soft}$ is equivalent to the position of $\tilde{u}=u+D$ in the frame of $\ket{C=D}_{soft}$. Since $\mathcal{A}$ and $\mathcal{A}_{D}$ are the same except for the difference of frames, they are connected by the following Ward identity (see Fig.~\ref{fig:tsca}).
\begin{equation}
\mathcal{A}=\mathcal{A}_{D}\,.
\label{WAR1}
\end{equation}
Here, since $\{\ket{C}\}_{soft}$ is a complete set of eigenstates of $\tilde{C}$, \eqref{WAR1} can be extended for any superposition state $\ket{\Psi}_{soft}$ as follows:
\begin{equation}
\mathcal{A}=\mathcal{A}_{\Psi}\,,
\label{WA2}
\end{equation}
where
\begin{align}
\mathcal{A}_{\Psi}=&\bra{\Psi}_{soft}\otimes\bra{0}_{hard}\phi(\tilde{u}_1,{z}_1)\cdots\phi(\tilde{u}_n,{z}_n)\nonumber\\
& \times\mathcal{S}\phi(\tilde{v}_{1},{z}_{n+1})\cdots\phi(\tilde{v}_m,{z}_{n+m})\ket{0}_{hard}\otimes\ket{\Psi}_{soft}\,.
\end{align}
Now, we note that off-diagonal components of the soft part vanish because of the antipodal matching condition. This is the expression of the Ward identity of supertranslation using dressed coordinates. 

\subsection{Factorization of soft graviton dynamics and dressing hard particles dynamics}

The equation \eqref{WA2} implies that the scattering amplitude using the dressed coordinates is independent of the soft sector of the state. This means that correlation functions of dressed particles $\phi(\tilde{u},z)$ are same no matter what the state of soft graviton $\ket{\Psi}_{soft}$ is. This reproduces the fact that the dynamics of particles produced by the dressed operator is independent of the motion of soft gravitons~\cite{Bousso:2017dny,Mirbabayi:2016axw} in a simpler way.

\subsection{Reproducing soft theorem}

Let us check that the Ward identity \eqref{WAR1} reproduces the soft theorem. Now, we can define an operator $\Pi$ conjugate to $\tilde{C}$. (see e.g. \cite{Bousso:2017dny})
\begin{equation}
\left[\Pi(z),\tilde{C}(w)\right] = i\delta(z-w)\,.
\end{equation}
It is known that $\Pi$ can be written in terms of Bondi news as follows:
\begin{equation}
\Pi = -\frac{1}{8\pi G}\gamma_{z{\bar{z}}} (D_{z}^2 N^{zz}_{\omega=0})\,,
\end{equation}
where $N^{zz}_{\omega=0}$ is zero frequency mode  of Bondi news. Therefore, $\ket{C=\epsilon}_{soft}$ can be constructed from $\ket{C=0}_{soft}$ as follows:
\begin{equation}
\ket{C=\epsilon}_{soft}=e^{i\int d^2z \epsilon \Pi}\ket{C=0}_{soft}
 =\left(1+i\int d^2z \epsilon \Pi\right)\ket{C=0}_{soft}+\mathcal{O}(\epsilon ^2)\,.\label{RE2}
\end{equation}
Also, 
\begin{equation}
 \phi(u+\tilde{C},z)=\phi(u,z)+\tilde{C}\partial_u\phi(u,z)+\mathcal{O}(\tilde{C}^2)\,.\label{RE1}
\end{equation}
Substituting these equations \eqref{RE2} \eqref{RE1}, $\epsilon=1/(z-w)$ and considering in the momentum space ($\partial_u\rightarrow iE_k$, $\partial_v\rightarrow -iE_k$), we obtain the soft theorem as follows
\begin{align}
&\bra{C=0}_{soft}\otimes\bra{0}_{hard}\phi(E_1,z_1)\cdots\phi(E_n,z_n)\nonumber\\
& \qquad\qquad \times :P_{w}\mathcal{S}:\phi(E_{n+1},z_{n+1})\cdots\phi(E_{n+m},z_{n+m}) \ket{0}_{hard}\otimes\ket{C=0}_{soft}\nonumber \\
&=\left[\sum_{k=1}^{m+n} \eta_k\frac{E_k}{w-z_k}\right] 
 \bra{C=0}_{soft}\otimes\bra{0}_{hard}\phi(E_1,z_1)\cdots\phi(E_n,z_n)\nonumber\\
&\qquad\qquad\qquad\qquad \times \mathcal{S}\phi(E_{n+1},z_{n+1})\cdots\phi(E_{n+m},z_{n+m})\ket{0}_{hard}\otimes\ket{C=0}_{soft}\,,
\end{align}
where $:\cdot:$ is time ordered product, $\eta_k=+1/-1$ if particle $k$ is outgoing/incoming and
\begin{equation}
P_w=\int d^2z \frac{\Pi}{z-w}\,.
\end{equation}
This equation is consistent with the result of \cite{Strominger:2013jfa, He:2014laa}.

\section{Hawking radiation from Black hole with soft hair}\label{HR}

In this section, we will see how Hawking radiation is modified when soft hair is taken into account. The introduction of dressed coordinates allows us to treat the problem following the standard description of the Unruh effect~\cite{Unruh:1976db}. For simplicity, we ignore backscattering, i.e., we consider the case where Hawking radiation originating in the vicinity of the horizon is freely propagating to ${\mathcal{I}}^+$ and the hair is ever-present rather than implanted dynamically. We also assume the adiabatic approximation, meaning that the evaporation process is slow enough such that at any time the static Schwarzschild spacetime is a good approximation. We also include a massless free scalar particle as matter.

\subsection{Near horizon dressed coordinate}

We begin by analyzing the near horizon region. The metric can be written as follows:
\begin{align}
\hat{g}&=-\hat{V}d\hat{v}_H^2+2d\hat{v}_Hd\hat{\rho}_H + \hat{r}^2\hat{\gamma}_{AB}d\hat{\Theta}_H^A\hat{\Theta}_H^B\nonumber\\
 &\sim -2\kappa\hat{\rho}_{H}d\hat{v}_H^2+2d\hat{v}_Hd\hat{\rho}_{H}+\hat{r}^2\hat{\gamma}_{AB}d\hat{\Theta}_H^Ad\hat{\Theta}_H^B\,.
\end{align}
We now introduce the tortoise coordinate $\hat{x}$ just inside the horizon as
\begin{equation}
 \hat{\rho}_H=-\frac{1}{2\kappa}e^{-2\kappa\hat{x}}\,.
\end{equation}

Then we have the following form of the metric.
\begin{equation}
 \hat{g}\sim e^{-2\kappa\hat{x}}(d\hat{v}_H^2+2d\hat{v}_Hd\hat{x})+4M^2\hat{\gamma}_{AB}d\hat{\Theta}_H^Ad\hat{\Theta}_H^B\,.
\end{equation}
In addition, we introduce the following retarded coordinate $\hat{u}_H$ just inside the horizon,
\begin{equation}
 \hat{u}_H=\hat{v}_H-2\hat{x}\,.
\end{equation}
We then obtain the following form of the metric.
\begin{equation}
 \hat{g}\sim-e^{2\kappa\hat{x}}d\hat{u}_Hd\hat{v}_H+4M^2\hat{\gamma}_{z\bar{z}}d\hat{z}_Hd\hat{\bar{z}}_H\,.
\end{equation}
Focusing on the vicinity of the north pole $|\hat{z}|<<1$, this metric becomes
\begin{equation}
  \hat{g}\sim-e^{2\kappa\hat{x}}d\hat{u}_Hd\hat{v}_H+4M^2d\hat{z}_Hd\bar{\hat{z}}_H
   =-e^{\kappa (\hat{v}_H-\hat{u}_H)}d\hat{u}_Hd\hat{v}_H+2d\hat{\zeta} d\bar{\hat{\zeta}}\,,
\end{equation}
where $\hat{\zeta}\equiv\sqrt{2}M\hat{z}_H$. If we take the plane approximation, the longitudinal mode becomes a plane wave solution. On the other hand, if we do not take the plane approximation, we can simply use spherical harmonic functions. In addition, by introducing coordinates for the free-falling observer,
\begin{equation}
 \hat{U}=-\frac{1}{\kappa}e^{-\kappa\hat{u}_H}\,, \quad
 \hat{V}=\frac{1}{\kappa}e^{\kappa\hat{v}_H}\,,
\end{equation}
the metric can be written as 
\begin{equation}
\hat{g}\sim -d\hat{U}d\hat{V}+2d\hat{\zeta} d\hat{\bar{\zeta}}\,,
\end{equation}
which represents the ordinary Minkowski spacetime.

\subsection{Thermofield double state for black hole}

In this section, we study Hawking radiation in the presence of soft hair. So far, as we have seen, this discussion is basically the same as the usual discussion of the Unruh effect in our notation using dressed coordinates. However, we will discuss it to clarify which quantities depend on soft gravitons. Also, since ingoing and outgoing modes are factorized,  we only consider the part of the wave that depends on the retarded coordinate.

\subsubsection{Dressed modes for free falling observer}

First, we construct the natural mode for a free-falling observer. Natural coordinates for free-falling observers are $\hat{U},\hat{V}$. Near the north pole of the horizon, the equation of motion in terms of these coordinates is 
\begin{equation}
\Box \phi = 0 \quad \Longleftrightarrow \quad
(-2\partial_{\hat{U}}\partial_{\hat{V}}+\partial_{\hat{\zeta}}\partial_{\hat{\bar{\zeta}}})\phi=0\,.
\end{equation}
Now, we turn our attention to modes that have finite frequencies when redshifted to be measured at the future null infinity. Therefore, we only need to focus on modes with high frequencies measured by the coordinate $\tilde{U}$, so we ignore derivatives along transverse direction,
\begin{equation}
\partial_{\hat{U}}\partial_{\hat{V}}\phi=0\,.
\end{equation}
We take the following as a dressed mode $\hat{f}_i$ that satisfies this equation of motion.
\begin{equation}
\hat{f}_i\propto e^{-i\omega_i\hat{U}}e^{ik\hat{\zeta}}e^{i\bar{k}\hat{\bar{\zeta}}}\,.
\end{equation}
The creation operator $\hat{a}^\dagger$ associated with this mode is defined as follows. 
\begin{equation}
\phi = \sum_i \hat{f}_i\hat{a}_i  + c.c.\,.
\end{equation}
This is essentially the same as the one introduced in a different way in e.g. \cite{Bousso:2017dny}. 

We consider the vacuum state for free-falling observers which is annihilated by the operators $\hat{a}_i$.

\subsubsection{Dressed modes for asymptotic observer}

Next, we construct modes for the asymptotic observer. The dressed coordinate made of observable quantities for the asymptotic observer is $\tilde{u}$. The equations of motion written in this coordinate are
\begin{align}
(-2\partial_{\tilde{u}}\partial_{\hat{v}_H}+e^{\kappa(\hat{v}_H-\tilde{u})}\partial_{\hat{\zeta}}\partial_{\bar{\hat{\zeta}}})\phi=0\,,
\end{align}
where $\hat{u}_{H}=\tilde{u}$. We are interested in the modes with finite frequencies as measured by the asymptotic observer. In this case, because of the blue-shift, taking the near horizon limit $\hat{v}_H-\tilde{u}\rightarrow 0$, only the first term remains. Thus, the equation of motion becomes
\begin{equation}
\partial_{\tilde{u}}\partial_{\hat{v}_H}\phi=0\,.
\end{equation}
Take the following as a dressed mode $\tilde{p}_i$ that satisfies this equation of motion. 
\begin{equation}
\tilde{p}_i\propto 
e^{-i\omega_i\tilde{u}}e^{ik\hat{\zeta}}e^{i\bar{k}\bar{\hat{\zeta}}}\,.
\end{equation}
We suppose that this mode has support only in the region where the coordinate $\tilde{u}$ is defined, i.e., outside the horizon. The creation operator $\tilde{b}_i^{\dagger}$ associated with this mode is defined as follows
\begin{equation}
\phi=\sum_i \tilde{p}_i \tilde{b}_i+c.c.\,.
\end{equation}

The region inside the horizon can be considered in the same way, defining the dressed mode $\hat{q}_i$ as follows. 
\begin{equation}
\hat{q}_i\propto 
e^{-i\omega_i\hat{u}_H}e^{ik\hat{\zeta}}e^{i\bar{k}\bar{\hat{\zeta}}}\,.
\end{equation}
This mode has support only inside the horizon. The creation operator $\hat{c}_i^{\dagger}$ is then introduced as follows.
\begin{equation}
\phi =\sum_i \hat{q}_i \hat{c}_i+c.c.\,.
\end{equation}

\subsubsection{Thermofield double state}

By combining these modes $\tilde{p}_i,\hat{q}_i$, we can create new modes for a free-falling observer as follows.
\begin{equation}
 \tilde{p}_i+e^{-\frac{\beta\omega_i}{2}}\bar{\hat{q}}_{\bar{i}}\,, \quad
\hat{q}_i+e^{-\frac{\beta\omega_i}{2}}\bar{\tilde{p}}_{\bar{i}}\,,
\end{equation}
where $\beta=\frac{2\pi}{\kappa}$ is the inverse temperature of the black hole, and the bar for the subscript indicates the parity transformation $\bar{i}=\{\omega_i, -k_i\}$. The following two sets of operators can be defined in relation to these modes.
\begin{equation}
 \tilde{b}_i-e^{-\frac{\beta\omega_i}{2}}\hat{c}^\dagger_{\bar{i}}\,, \quad
\hat{c}_i-e^{-\frac{\beta\omega_i}{2}}\tilde{b}^\dagger_{\bar{i}}\,.
\end{equation}
The state we are now considering is the vacuum for a free-falling observer that is annihilated by these two sets of annihilation operators. Such a state can be written as a soft graviton-dependent thermofield double (TFD) state as follows 
\begin{equation}
\ket{TFD}=\sum_{C,C_H}g(C,C_H)\frac{\exp\left(\sum_i e^{-\frac{-\beta\omega_i}{2}}\hat{c}^\dagger_{\bar{i}}(C_H)\tilde{b}^\dagger_i(C)\right)}{\sqrt{Z}}\ket{0}^{hard}_{\mathcal{H}}\ket{0}^{hard}_{{\mathcal{I}}^+}\otimes\ket{C_H}^{soft}_{\mathcal{H}}\ket{C}^{soft}_{{\mathcal{I}}^+}\,.
\label{TFD}
\end{equation}
The dependence of the soft hair state on the c-number of soft hair operators is made explicit by expanding them in the eigenstate basis as
\begin{equation}
\ket{\psi}_{soft}=\sum_{C,C_H}g(C_H,C)\ket{C_H}^{soft}_{\mathcal{H}}\ket{C}_{{\mathcal{I}}^+}^{soft}\,.
\end{equation} 
Tracing out the black hole degrees of freedom, we then obtain the density matrix of Hawking radiation
\begin{equation}
\rho=\sum_{C} P(C) \tilde{\rho}_{th}(C) \otimes\ket{C}_{soft}\bra{C}_{soft}\,,
\end{equation}
where $P(C)$ is the probability of asymptotic soft hair $\tilde{C}$ observed as $C$, and 
\begin{equation}
   \tilde{\rho}_{th} = \frac{e^{-\beta \tilde{H}}} {Z}=\bigotimes_i\sum_{n=0}^{\infty}
\frac{e^{-\beta n \omega_i}}{Z_i}\ket{\tilde{n}_i}_{hard}\bra{\tilde{n}_i}_{hard}\,.
\end{equation}
Here, $\tilde{H}$ is the Hamiltonian,
\begin{equation}
   \tilde{H} = \sum_i \omega_i \tilde{b}_i^\dagger \tilde{b}_i\,,
\end{equation}
and $Z$ and $Z_i$ are partition functions of free bosons,
\begin{equation}
    Z=\prod_i Z_i\,, \quad Z_i=\frac{1}{1-e^{\beta\omega_i}}\,.
\end{equation}

\subsubsection{Changing Hawking radiation by soft gravitons in asymptotic region}

Now we see how Hawking radiation is modified by soft graviton in the asymptotic region. To do so, we would like to focus on undressed modes and creation operators that correspond to observers associated with the $u$ coordinate,
\begin{equation}
\phi(u,z) =\sum_j b_j p_j +c.c.\,.
\end{equation}
To describe the Hawking radiation as measured by the observer, the relationship between dressed and undressed quantities can be written in terms of soft graviton dependent Bogoliubov coefficients as
\begin{align}
\tilde{p}_i&=\sum_j \tilde{A}_{ij} p_j\,,
\end{align}
where
\begin{align}
\tilde{A}_{ij} &= \int d\zeta d\bar{\zeta} \exp
\left(ik_i\zeta -ik_j\zeta+i\omega_i \tilde{C}+c.c.\right)\delta(\omega_j-\omega_i)\,.
\end{align}
Here, the Bogoliubov coefficient is evaluated at ${\mathcal{I}}^+$, $v\rightarrow\infty$ as it is independent of the Cauchy surface on which it is evaluated.

The relationship between the creation operators can be written in terms of the Bogoliubov coefficients as 
\begin{align}
b_j&=\sum_i \tilde{A}_{ij} \tilde{b}_i\,.
\end{align}
From this, it is possible to rewrite the Hamiltonian $\tilde{H}$ that characterizes the mixed state representing Hawking radiation as follows;
\begin{equation}
\tilde{H}=\sum_{ijk} \omega_j \tilde{A}_{ji} \tilde{A}^\ast_{jk}b_i^\dagger b_k\,.
\end{equation}
Therefore, it is clear that as long as we observe the dressed mode $\tilde{b}$, the Hawking radiation is unaltered. This is consistent with the assertion of Refs.~\cite{Bousso:2017dny,Mirbabayi:2016axw} that the dynamics of the dressed mode do not depend on the state of soft gravitons.

\section{Discussion}\label{DIS}

In this paper, we have introduced the notion of dressed coordinates and demonstrated the relevance of such a definition by reproducing some known results in the literature in a simplified way. We then computed Hawking radiation for a black hole with soft hair and showed that the physical observables are unchanged if we consider the dressed modes. We could therefore be tempted to conclude that soft modes cannot have any relevance for the resolution of the information loss paradox. 

However, this conclusion might change if the hair is implanted dynamically, for example, by the backreaction of Hawking radiation. To develop a physical intuition, let us consider the memory effect. The memory effect consists of a permanent displacement due to the passage of a gravitational perturbation~\cite{Favata:2010zu}. If we consider a detector in flat spacetime, before and after the passage of gravitational waves the metric is described by the flat Minkowski metric written in different coordinates. It is therefore impossible to detect the memory effect unless the detector is set up before the passage of the gravitational waves. This would fix the gauge and we would no longer be allowed to change coordinates, making the memory effect observable at least in principle. 

\begin{figure}[htbp]
\begin{center}
\includegraphics[trim=40 40 30 40,width=70mm]{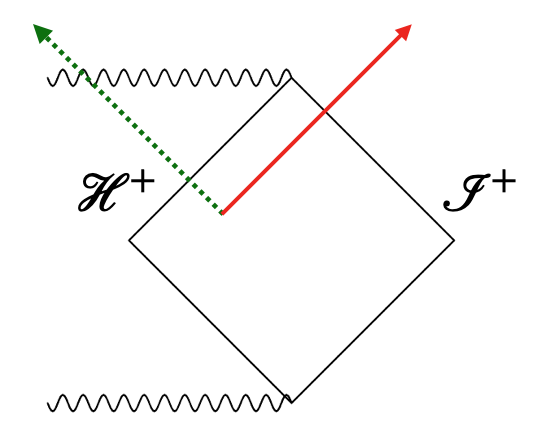}
\caption{The green dashed arrow indicates negative energy flux which flows into the horizon. And the red solid arrow indicates positive energy flux which flows out towards the future null infinity. It is expected that horizon soft hair and soft gravitons in asymptotic region will be affected by their backreaction. 
}
\label{FMA2}
\end{center}
\end{figure}

For the soft hair to be relevant, something similar should happen. As we have shown, we can always remove the effect of soft hair unless we fix the detector before implanting the hair. To study this possibility, we would need to investigate how Hawking radiation can implant soft hair. Specifically, we can expect soft hair (soft gravitons in the asymptotic region) to change as the energy flux passes through the horizon (asymptotic region), see Figure.\ref{FMA2}.

The following quantum gravitational effects, which cannot be taken into account by merely including a semiclassical backreaction, are necessary to contribute to the information paradox. First, the information of the Hawking partner needs to be transferred to soft hair. This is not just the entanglement of Hawking partners with soft hair. An outside observer must be able to regard the Hawking partner's degrees of freedom as soft hair degrees of freedom. If we think of soft hair as the origin of Bekenstein-Hawking entropy, this argument must be correct. Second, soft hairs which have Hawking partner's information must be returned to the asymptotic region. The Hawking process is inevitable if we require that the vicinity of the horizon be a vacuum for the falling observer. Also, if unitarity is required, the entanglement partner of the  early Hawking radiation must be returned to the asymptotic region to purify the early Hawking radiation. From these two requirements, in addition to the Hawking process, the process of soft hair returning to the asymptotic region is necessary. This new process inevitably violates the laws of semiclassical general relativity~\footnote{In the context of quantum teleportation~\cite{Bennett:1992tv}, the amount of such violation may in principle be significantly reduced~\cite{Mukohyama:1998xq}. However, the transfer of the classical channel still requires violation of the laws of semiclassical general relativity.}. If this occurs in the time scale known as Page time~\cite{Page:1993df,Page:1993wv} then it may correspond to the formation of islands in the island prescription~\cite{Penington:2019kki,Almheiri:2019qdq}. On the other hand, if semiclassical general relativity is a valid description of the system before reaching the final stage of the black hole evaporation then the information returning process may be possible only in the Planckian regime~\cite{Buoninfante:2021ijy}. It is important to study this issue to see if which is the case, taking into account the backreaction and the quantum gravity effects. 

The study of the effects of the backreaction that we leave for future investigation are expected to be technically difficult in 4 spacetime dimensions. It might be possible to extract some information about them considering a simplified setup of 2-dimensional gravity such as the Jackiw--Teitelboim (JT) theory of gravity. Soft hair is a phenomenon in which non-physical degrees of freedom that should be gauge redundancy become physical (edge modes) due to the existence of a boundary on the Cauchy surface when a null hypersurface such as a horizon or a null infinity in an asymptotically flat spacetime is taken as (a portion of) a Cauchy surface. The simplification is possible because edge modes also exist in JT gravity and actually there are only edge mode degrees of freedom due to the absence of gravitational waves. 

As a final comment, let us note that the non-commutative nature of the soft-hair operators (supertranslation charge) did not show up in the concrete calculations in the present paper. Hence there was virtually no difference between classical and quantum theory due to the upgrading of the coordinates to operators. On the other hand, in future works we hope to incorporate the backreaction of the Hawking radiation to spacetime (and hopefully the quantum gravity effects discussed above). In this case, the dressed coordinates and non-commutative operators introduced in this paper may play nontrivial roles. We then hope that this may allow us to see if soft hair can provide clues toward the resolution of the information loss paradox as proposed by Hawking \textit{et al.}~\cite{Hawking:2016msc,Hawking:2016sgy,Hawking:2015qqa}.

\section*{Acknowledgements}
We thank very much Massimo Porrati and Tadashi Takayanagi for valuable comments. FDF acknowledges financial support by Japan Society for the Promotion of Science Grants-in-Aid for international research fellow No. 21P21318. The work of SM was supported in part by World Premier International Research Center Initiative, MEXT, Japan. The work of NO and TW was supported by JST, the establishment of university fellowships towards the creation of science technology innovation, Grant Number JPMJFS2123.

\bibliography{soft} 
\bibliographystyle{utphys}
\expandafter\ifx\csname ifdraft\endcsname\relax

\end{document}